\documentclass[%
reprint,
superscriptaddress,
 amsmath,amssymb,
 aps,
pra
]{revtex4-2}

\usepackage{graphicx}
\usepackage{dcolumn}
\usepackage{bm}
\usepackage{dsfont}
\usepackage{physics}
\usepackage[table]{xcolor}
\usepackage{enumitem}
\usepackage{tikz}
\usepackage{circuitikz}
\usepackage{verbatim}
\usepackage[normalem]{ulem}
\usepackage[colorlinks]{hyperref}

\usepackage{newtxtext}
\usepackage{newtxmath}
\usepackage[cal=cm]{mathalfa}

\newcommand{\A}{\mathsf{A}}

\newcommand{\M}{\vvmathbb{M}}
\newcommand{\Mdot}{\dot{\vvmathbb{M}}}
\newcommand{\HSM}{H_{SM}}

\newcommand{\wi}{w_\qu[\rho_i]}
\newcommand{\dFtilde}{\Delta \tilde{F}}
\newcommand{\dF}{\Delta F}
\newcommand{\zz}{\mathbf{z}}
\newcommand{\ensavg}{\vvmathbb{E}}
\newcommand{\rhoA}{\rho_A}
\newcommand{\rhoB}{\rho_B}
\newcommand{\cl}{\text{cl}}
\newcommand{\qu}{\text{qu}}
\newcommand{\K}{\mathcal{K}}

\newcommand{\id}{\ensuremath{\mathds{1}}}
\newcommand{\kb}[2]{|#1\rangle\langle#2|}

\newcommand{\diff}{\mathrm{d}}

\newcommand{\stkout}[1]{\ifmmode\text{\sout{\ensuremath{#1}}}\else\sout{#1}\fi}

\makeatletter
\def\maketitle{
\@author@finish
\title@column\titleblock@produce
\suppressfloats[t]}
\makeatother

\begin{document}

\title{Operational work fluctuation theorem for open quantum systems}




\author{Konstantin Beyer}
\affiliation{Department of Physics, Stevens Institute of Technology, Hoboken, New Jersey 07030, USA}
\affiliation{Institute of Theoretical Physics, TUD Dresden University of Technology, 01062, Dresden, Germany}
\author{Walter T. Strunz}
\affiliation{Institute of Theoretical Physics, TUD Dresden University of Technology, 01062, Dresden, Germany}

\date{\today}

\begin{abstract}
The classical Jarzynski equality establishes an exact relation between the stochastic work performed on a system driven out of thermal equilibrium and the free energy difference in a corresponding quasi-static process. 
This {fluctuation theorem} bears experimental {{relevance}}, as it enables the determination of  the free energy difference through the measurement of externally applied work in a nonequilibrium process.
In the quantum case, the Jarzynski equality only holds if the measurement procedure of the stochastic work is drastically changed{{: it is}} replaced by a so-called two-point measurement (TPM) scheme {{that}}  requires the knowledge of the initial and final Hamiltonian {{and therefore}}  {{lacks the predictive power}} for the free energy difference that the classical Jarzynski equation is known for. 
Here, we propose a quantum {{fluctuation theorem}} that is valid for externally measurable quantum work determined during the driving protocol. In contrast to the TPM case, {the theorem also applies to open quantum systems and the scenario can be realized without knowing the system Hamiltonian.}
{{Our fluctuation theorem}} comes in the form of an inequality and therefore only yields bounds to the true free energy difference. The inequality is saturated in the quasiclassical case of vanishing energy coherences at the beginning and at the end of the protocol. Thus, there is a clear quantum disadvantage. 

\end{abstract}

\maketitle

\paragraph*{Introduction---}

One of the central fluctuation theorems in nonequilibrium thermodynamics is the so-called Jarzynski equality (JE)~\cite{jarzynskiEquilibriumFreeenergyDifferences1997,jarzynskiNonequilibriumEqualityFree1997}
\begin{align}
\label{eq:JE-general}
    \ensavg[e^{-\beta w}] = e^{-\beta \dF}.
\end{align}
Here, \(\ensavg\) denotes the ensemble average over trajectories of an initially thermalized system that is driven out of equilibrium, \(w\) is the work performed along such a trajectory, \(\beta\) is the inverse temperature, and \(\dF\) is the free energy difference between Gibbs states corresponding to the initial and the final Hamiltonian of the driving process. 
The relevance of this equation comes from the fact that it establishes an exact relation between the work determined in a  nonequilibrium process on the left-hand side and the quasistatic quantity \(\dF\) on the right-hand side. 

Originally derived for classical systems, the JE was soon shown to be valid for quantum systems as well.
However, to make sense of Eq.~\eqref{eq:JE-general}, it is crucial to specify how the work \(w\)  is actually determined, in particular since the commonly adopted definition of \(w\) differs considerably between the classical and the quantum case. 

In the classical scenario, it is usually assumed that the work can be measured as an integral of power over the duration of the driving process~\cite{jarzynskiWorkFluctuationTheorems2006,hummerFreeEnergyReconstruction2001,hummerFreeEnergySurfaces2005,hummerFreeEnergyProfiles2010}. In experiments, this quantity can be determined by externally measuring the force that drives the system out of equilibrium and integrating it over the displacement, \(w_\cl = \int F \diff \lambda\)~\cite{liphardtEquilibriumInformationNonequilibrium2002,harrisExperimentalFreeEnergy2007,ramanDecipheringScalingSinglemolecule2014,walhornExploringSulfataseCatch2018,sanchezDeterminationProteinProtein2022,preinerFreeEnergyMembrane2007,mossaDynamicForceSpectroscopy2009,junierRecoveryFreeEnergy2009,tangUnderwaterAdhesionStimuliResponsive2018,houExtractMotiveEnergy2022}. No knowledge of the underlying system Hamiltonian is required.

\begin{table}
\begin{tabular}{l|c|c}
     & \begin{tabular}{@{}c@{}}Externally measurable \\ during the protocol, \\ unknown Hamiltonian, \\ open system\end{tabular}& \begin{tabular}{@{}c@{}}TPM scheme measured\\ on the system, \\ known Hamiltonian, \\ closed system\end{tabular} \\ \hline
    Cl. & \cellcolor{green!25} $ \phantom{\Bigg(}\ensavg[e^{-\beta w_\cl}] = e^{-\beta \dF} \phantom{\Bigg)}$ & \cellcolor{cyan!25} $\phantom{\Bigg(}\ensavg[e^{-\beta w_\text{TPM}}] = e^{-\beta \dF}\phantom{\Bigg)}$ \\ \hline
    Qu. &  \cellcolor{yellow!25} ${\phantom{\Bigg(} \ensavg[e^{-\beta  {w_\qu}}] \leq e^{-\beta \dF}\phantom{\Bigg)}}$ & \cellcolor{cyan!25} $\phantom{\Bigg(}\ensavg[e^{-\beta w_\text{TPM}}] = e^{-\beta \dF} \phantom{\Bigg)}$ \\ \hline
    \end{tabular}
    \caption{In both the classical and quantum cases, work can be defined either by an integrated work measurement during the protocol or by a TPM scheme, the latter being restricted to closed system dynamics. For classical systems, both definitions satisfy the Jarzynski equality. In the quantum case, usually only the TPM scheme is considered, which also satisfies the identity. We address the missing piece of a quantum scenario with work as a quantum expectation value that is externally measurable during the protocol, and show that the fluctuation theorem becomes an inequality, providing only an upper bound for the true free energy difference \(\dF\). This quantum disadvantage is a direct consequence of the fact that the quantum density operator of a Gibbs state does not describe an \emph{objective} ensemble of orthogonal energy eigenstates.}
    \label{tab:table}
\end{table}

The quantum JE, by contrast, is commonly based on the so-called two-point measurement (TPM) scheme~\cite{tasakiJarzynskiRelationsQuantum2000,kurchanQuantumFluctuationTheorem2001,talknerFluctuationTheoremsWork2007,engelJarzynskiEquationSimple2007,campisiColloquiumQuantumFluctuation2011,espositoNonequilibriumFluctuationsFluctuation2009,huberEmployingTrappedCold2008,perarnau-llobetNoGoTheoremCharacterization2017,hovhannisyanEnergyConservationJarzynski2021,bartolottaJarzynskiEqualityDriven2018,mohammadySelfconsistencyTwopointEnergy2021}. A projective measurement of the system Hamiltonian is performed in the beginning (\(H_0\)) and at the end (\(H_T\)) of the driving process, and---assuming that there is no heat transfer---the work is defined as the difference between the final and the initial energy eigenvalue outcome, \(w_\text{TPM} = E_T - E_0\). 
Experimental tests of the quantum JE based on the TPM scheme have been reported for various platforms such as trapped ions, superconducting qubits, and nuclear spins~\cite{batalhaoExperimentalReconstructionWork2014,anExperimentalTestQuantum2015,cerisolaUsingQuantumWork2017,zhangExperimentalDemonstrationWork2018,hahnQuantumManyBodyJarzynski2023}.

It is important to notice that, despite being formally equivalent, the quantum JE is not the analog of the original classical one, but should rather be compared to a TPM version of the classical scenario.
The situation is summarized in Tab.~\ref{tab:table}. 
The upper left and the lower right corner represent the standard cases considered in classical and quantum scenarios, respectively.
Applying the TPM scheme to a classical system also fulfills the JE (upper right corner). 
This case, however, has no relevance in the literature, which comes as no surprise. In comparison to the original classical JE scenario with externally measurable work, the TPM version---be it classical or quantum---has two serious drawbacks.

(i) The TPM scheme is valid only for a closed system without heat exchange, as work equals the change in internal energy only in this case. In contrast, the original scenario considered by Jarzynski also applies to systems that partially thermalize during the driving protocol, thus bridging the gap between the limiting cases of a rapidly changing Hamiltonian with (almost) no heat transfer and the quasistatic scenario where the system is in thermal equilibrium at all times. There are extensions of the TPM scheme to cover the thermalizing case~\cite{ramezaniQuantumDetailedBalance2018,naghilooHeatWorkIndividual2020,scandiQuantumWorkStatistics2020,strasbergFirstSecondLaw2021}. However, as far as we can see, they all depend on details of how the heat transfer is modeled and measured and thus cannot match the elegance of the original JE, which only requires that the (partial) thermalization preserves the Gibbs state of the instantaneous Hamiltonian~\cite{jarzynskiEquilibriumFreeenergyDifferences1997}.

(ii) To the best of our knowledge, all TPM schemes considered in the literature require the system Hamiltonians \(H_0\) and \(H_T\) to be known in advance in order to perform the energy measurements. 
On the one hand, this is impractical for all but the smallest systems like qubits. {(Consider the stretched molecule in Fig.~\ref{fig:molecule}
, where even classically, the full Hamiltonian is inaccessible, and measuring the microstate's energy would require determining the position and momentum of each atom.)}
On the other hand---and we consider this to be conceptually the main problem of any TPM JE---the knowledge of the system Hamiltonian {{turns the quantum}} JE into a mathematical identity with {{no predictive power}} since the free energy difference \(\dF\) on the right-hand side of Eq.~\eqref{eq:JE-general} can be calculated directly from the {{known}} Hamiltonians without even performing the experiment. 
We note that ancilla-assisted quantum work measurements (see Refs.~\cite{mazzolaMeasuringCharacteristicFunction2013,cerisolaUsingQuantumWork2017,dornerExtractingQuantumWork2013,roncagliaWorkMeasurementGeneralized2014,rubinoQuantumSuperpositionThermodynamic2021,rubinoInferringWorkQuantum2022}) do not resolve this issue, but only shift it to the choice of the correct coupling Hamiltonian between system and ancilla.

Motivated by the observations (i) and (ii), we address the missing lower left corner in Tab.~\ref{tab:table}. This will be the main result of this letter: a quantum fluctuation theorem of Jarzynski form based on an externally measurable work definition without knowledge of the system Hamiltonian, which also works for open systems.
As we will show, the fluctuation theorem is no longer {{an equality}} in this case, but the best we can get is an inequality giving bounds on the true free energy difference \(\dF\). Interestingly, quasiclassical cases without energy coherences saturate the bound, demonstrating a clear quantum disadvantage compared to classical systems.

In the remainder of this letter, we first highlight some important features of the classical case. Based on this analysis, we present a quantum analog and state our theorem. 
 {We first address the special case of a closed quantum system and its experimental relevance. We then extend the scenario to the general case of an open quantum system and conclude.}

\begin{figure}
    \centering
    \includegraphics[width=.6\linewidth]{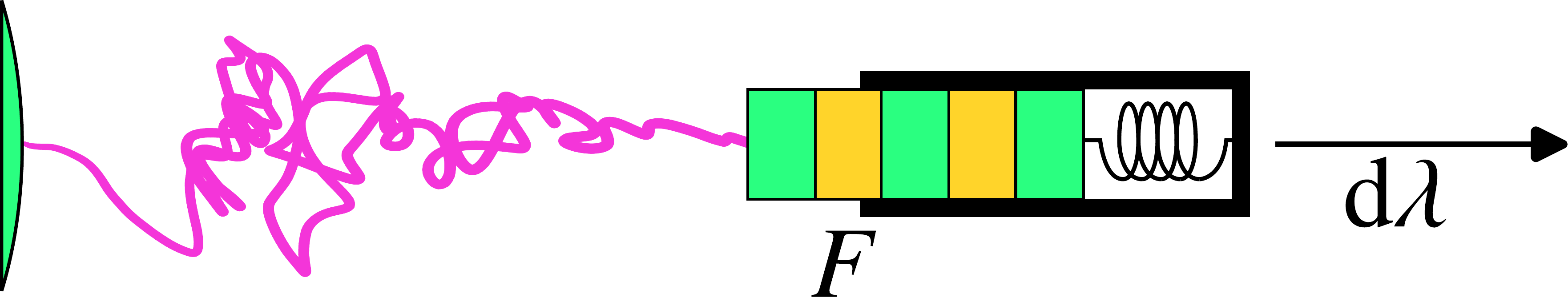}
    \caption{Scheme of a single-molecule force spectroscopy experiment. The atom is stretched by moving the rest position \(\lambda\) of the spring scale. The applied work, which depends on the initial microstate of the molecule, can be determined as an integral of the measured force \(F\) over the displacement \(\diff \lambda\). Experiments that use these work measurements to calculate free energy differences through the Jarzysnki equality have been successfully implemented~\cite{liphardtEquilibriumInformationNonequilibrium2002,harrisExperimentalFreeEnergy2007,ramanDecipheringScalingSinglemolecule2014,walhornExploringSulfataseCatch2018,sanchezDeterminationProteinProtein2022,preinerFreeEnergyMembrane2007,mossaDynamicForceSpectroscopy2009,junierRecoveryFreeEnergy2009,tangUnderwaterAdhesionStimuliResponsive2018,houExtractMotiveEnergy2022}. Importantly, these experiments are ignorant about the Hamiltonian of the molecule.}
    \label{fig:molecule}
\end{figure}

\paragraph*{Classical case---}
The prototypical classical experiment is depicted in Fig.~\ref{fig:molecule}~\cite{jarzynskiWorkFluctuationTheorems2006,harrisExperimentalFreeEnergy2007}. 
The variation of the system Hamiltonian \(H_\lambda\) emerges from the change of an external parameter \(\lambda\), in this case the rest position of the spring scale to which the molecule is attached. 
During the change of the displacement \(\lambda\), the force \(F\) applied to the molecule can be measured, and the work is given by \(w_\cl = \int_{0}^T F \dot\lambda \, \diff t\). 

Initially, the system is in a thermalized Gibbs state \(\rho_0\). 
However, let us assume for the sake of conceptual clarity that the system is closed during the driving process (open systems are discussed below). 
In each run, an initial microstate \(\zz_0\) is probabilistically singled out when the protocol starts, but the remaining trajectory is fully deterministic in this case without heat exchange. 
Accordingly, the measured work is equal to the energy difference
\begin{align}
\label{eq:w-classical}
	w_\cl[\zz_0] = H_T[\zz_T] - H_0[\zz_0],
\end{align}
which only depends on the initial microstate \(\zz_0\) since the final one \(\zz_T\) follows deterministically. 
The quantity \(w_\cl[\zz_0]\) fulfills the JE~\cite{jarzynskiEquilibriumFreeenergyDifferences1997}
\begin{align}
\label{eq:JE-classical}
    \ensavg[e^{-\beta w_\cl}] =\int \diff\zz_0\, \rho_0[\zz_0]\, e^{-\beta w_\cl[\zz_0]} = e^{-\beta \dF}.
\end{align}

Two points are crucial for the external measurement of work, as shown in Fig.~\ref{fig:molecule}, to be usable in the classical JE.
\begin{enumerate}[label=(\Roman*),wide, labelindent=0pt]
    \item \label{it:1} Although the initial microstate \(\zz_0\) remains hidden to the experimenter, there \emph{is} an objective microstate in the classical system in each run.  
    \item \label{it:2} For each possible trajectory in the system, the work \(w_\cl\) can be measured with certainty in a single run. 
\end{enumerate} 
This allows the integral over an unknown microstate \(\zz_0\) in Eq.~\eqref{eq:JE-classical} to be operationally replaced by the ensemble average \(\ensavg\) over the measured values of work \(w_\cl\), enabling experiments like the one shown in Fig.~\ref{fig:molecule}. (We note that there are various subtleties for the implementation of the classical experiment, but those are of no conceptual relevance here~\cite{talknerColloquiumStatisticalMechanics2020,shirtsEquilibriumFreeEnergies2003,jarzynskiRareEventsConvergence2006,hummerFreeEnergyProfiles2010,dengDeformedJarzynskiEquality2017}.)   

In classical systems, these requirements are so obviously fulfilled that they usually go unstated. However, they provide valuable insights when considering quantum systems, where these conditions are just as obviously not met.
Unlike \ref{it:1} for classical systems, a quantum system is never in an objective state unless we specify by what measurement we have obtained this information.
Regarding condition \ref{it:2}, textbook quantum mechanics teaches us that physical quantities can generally only meaningfully be measured as expectation values and not in single experimental runs. 
Keeping these facts in mind, we now propose a quantum analog of this classical scenario.

\paragraph*{Quantum analog---}
 {Let us start with the special case of a closed quantum system.}
Consider an apparatus that does two things. 
\begin{enumerate}[wide]
    \item \label{apparatus-1}  It effectively implements a parameter-dependent Hamiltonian \(H_\mu\) on a closed quantum system, thereby transforming it by a unitary \(U\). The experimenter has control over the parameter \(\mu\) but does not know which Hamiltonian is actually implemented for a given \(\mu\) (analogous to \(\lambda\) in the classical example in Fig.~\ref{fig:molecule}).
    \item \label{apparatus-2}  The apparatus can measure the work applied to the quantum system when \(\mu\) is continuously changed (as in the classical case). However, this work is obtained as an expectation value. That is, for a fixed protocol \(\mu_t\) with initial \(\mu_0\) and final \(\mu_T\) and an arbitrary {{(unknown)}} input state \(\rho\), the apparatus yields outcomes that average to the following expectation value after sufficiently many runs:
\begin{align}
\label{eq:expected-work} \nonumber
     {w_\qu[\rho]} & = \tr[\rho(T)H_{T}] - \tr[\rho H_{0}] \\
    & = \tr\left[\rho\! \left(U^\dag H_{T} U - H_{0}\right)\right],
\end{align}
where \(H_0 = H_{\mu_0}\), \(H_T = H_{\mu_T}\), and \(U\) is the unknown unitary dynamics generated by the time-dependent system Hamiltonian \(H_{\mu_t}\) {{up to time \(T\)}}.
\end{enumerate} 

 {The concrete implementation of such an apparatus is not crucial for the results of this letter. A possible construction that does the job to arbitrary precision was presented in Refs.~\cite{beyerMeasurementQuantumWork2023,beyerWorkExternalQuantum2020}}
We give a concise review of it in App.~\ref{app:work-measurement}.
 {Such an apparatus} is the quantum analog of the externally measured work in Fig.~\ref{fig:molecule} that operationally also only depends on the fixed protocol of the parameter \(\lambda_t\) but is otherwise agnostic about the involved Hamiltonians and the microstate  \(\zz_0\).
The crucial difference to the classical case is the fact that  {in} quantum  {physics, in general, measurement outcomes} of a single run  {are meaningless.}  {Instead, expectation values have to be measured}. 

Looking at an initially thermalized system as in the Jarzynski experiment, this means that our apparatus can measure the average work  {\(W_\text{avg} = w_\qu[\rho_0]\)} applied to the Gibbs state \(\rho_0\) during the driving given by the protocol \(\mu_t\).
From the second law of thermodynamics we know that this quantity provides a bound to the free energy difference,
\(W_\text{avg} \geq \dF\).
Importantly, with the apparatus just described, this bound \(W_\text{avg}\) is operationally accessible without the knowledge of the implemented Hamiltonian \(H_{\mu_t}\) (see Fig.~\ref{fig:quantum-scheme}\,a).
However, we can do better by invoking the conditions \ref{it:1} and \ref{it:2} that enable the classical JE. 

In order to measure the work applied to single microstates as in the classical case, we have to objectify these states by quantum measurements (to overcome the contradiction of quantum mechanics with \ref{it:1}), and the work applied to such an objective initial state has to be determined in multiple runs (to overcome the contradiction with \ref{it:2}). 
\begin{figure}
    \centering
    \includegraphics[width=1.\linewidth]{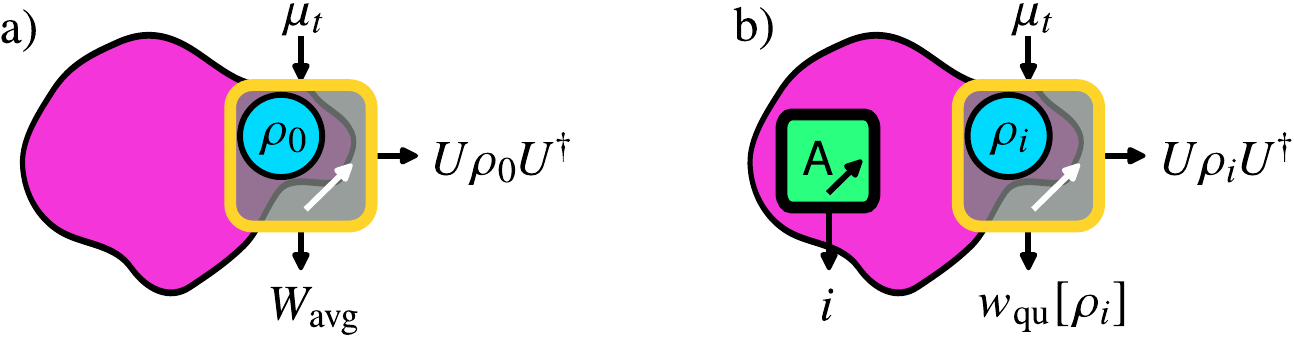}
    \caption{Scheme of the quantum scenario. The initial thermal state \(\rho_0\) is the partial trace of a larger joint state with the environment. The apparatus implements a protocol \(\mu_t\), leading to a time-dependent system Hamiltonian and a resulting unitary evolution \(U\).    
    a) In multiple runs, the apparatus can determine the expectation value of the average work \(W_\text{avg} {=w_\qu[\rho_0]}\). This quantity provides an upper bound for the free energy difference \(\dF\). b) By measurements on the environment, the thermal state in the system can be decomposed into an ensemble \(\mathcal{D} = \{(p_i,\rho_i\}\). Average work values   {\(w_\qu[\rho_i]\)} conditioned on the outcomes \(i\) fulfill a fluctuation theorem which always provides a tighter bound \(\dFtilde\), i.e., \(W_\text{avg} \ge \dFtilde \ge \dF\).  }
    \label{fig:quantum-scheme}
\end{figure}
This is achieved by the following scenario (see  Fig.~\ref{fig:quantum-scheme}\,b).

The thermal state \(\rho_0\) of the system  always emerges as the partial trace of a larger joint state of system and environment, i.e., \(\rho_0=\tr_E[\rho_{SE}]\).
Now assume the experimenter performs a measurement on the environment given by a positive operator-valued measure (POVM) \(\A=\{\A_i\}\), with positive operators \(\A_i\) and \(\sum_i \A_i = \id\).
Depending on the outcome \(i\), the state of the system is given by \(\rho_i = p_i^{-1}\tr_E[\rho_{SE}(\id_S\otimes \A_i)]\), 
with \(p_i = \tr[\rho_{SE} (\id\otimes\A_i)]\).
We call the set of tuples \(\mathcal{D} = \{(p_i,\rho_i)\}\) a \emph{decomposition} of the Gibbs state \(\rho_0\) since by construction \(\sum_i p_i \rho_i = \rho_0\). 

 {For a macroscopic environment, it seems unlikely that an indirect measurement can reveal any information about the system state and the decomposition \(\mathcal{D}\) would be trivial in that case. However, in a typical microscopic quantum thermodynamic setup the environment is often well controllable. 
Experiments that are capable of indirect measurements have been described and implemented on various platforms including trapped ions, cavity QED, superconducting circuits, and nitrogen-vacancy centers~\cite{horowitzQuantumtrajectoryApproachStochastic2012,hekkingQuantumJumpApproach2013,gongQuantumtrajectoryThermodynamicsDiscrete2016,maslennikovQuantumAbsorptionRefrigerator2019,stahlDemonstrationEnergyExtraction2024,najera-santosAutonomousMaxwellsDemon2020,vonlindenfelsSpinHeatEngine2019,naghilooHeatWorkIndividual2020,jiSpinQuantumHeat2022}.}

Note that the measurement \(\A\) must not be performed on the system itself, as without knowing the initial energy eigenbasis almost all measurements would disturb the initial state \(\rho_0\) on average, thereby injecting energy to the system (this effect is employed in so-called measurement-fueled engines~\cite{mohammadyQuantumSzilardEngine2017,yiSingletemperatureQuantumEngine2017,elouardExtractingWorkQuantum2017,seahMaxwellLesserDemon2020,dasMeasurementBasedQuantum2019,ankaMeasurementbasedQuantumHeat2021,opatrnyWorkGenerationThermal2021,manikandanEfficientlyFuelingQuantum2022}).

We assume that the experimenter has no knowledge about the joint state \(\rho_{SE}\) other than the fact that it locally reduces to the thermal state \(\rho_0\) of the system. We furthermore assume that they do not even know what POVM \(\A\) is implemented by the measurement on the environment. Therefore, the experimenter cannot know the conditional state \(\rho_i\). The only information obtained by the measurement is the label \(i\) of the outcome.
This again is in strong analogy to the classical case. As stated in condition~\ref{it:1}, also there, the experimenter does not know the microstate \(\zz_0\) but only relies on the fact that there \emph{is} an objective state.

After the measurement on the environment, the experimenter runs the protocol \(\mu_t\) and measures the work. 
Repeating this procedure many times, the experimenter can now condition the expectation value of the work measurement on the outcomes \(i\) of the initial measurement on the environment to obtain
\begin{align}\label{work_i}
    {\wi} = \tr[\rho_i(U^\dag H_T U- H_0)].
\end{align}
The measured \(i\)-dependent work satisfies an inequality.

\paragraph*{Fluctuation theorem---}
Consider a quantum system in a Gibbs state \(\rho_0\) at inverse temperature \(\beta\) with respect to a Hamiltonian \(H_0\), i.e., \(\rho_0 = Z_0^{-1} \exp[-\beta H_0]\) with \(Z_0=\tr \exp[-\beta H_0]\). 
For any decompositions \(\mathcal{D} = \{(p_i,\rho_i)\}\) of \(\rho_0\), any Hamiltonian \(H_T\), and any unitary \(U\) the following inequality holds for the 
 {\(w_\qu[\rho_i]\)} from Eq.~\eqref{work_i}:
\begin{align}
\label{eq:theorem}
   \ensavg[e^{-\beta   {w_\qu}}] \coloneq 
   \sum_i p_i e^{-\beta   {\wi} } \leq  e^{-\beta \dF},
\end{align}
where \(\dF = -\beta^{-1}\ln(Z_T/Z_0) \) with \(Z_T = \tr\exp[-\beta H_T]\). 

The crucial point and distinctive feature of this inequality is the fact that the decomposition \(\mathcal{D}\) can involve non-orthogonal states and, in particular, is in general not the eigendecomposition of the Gibbs state \(\rho_0\), in contrast to the case considered in Refs.~\cite{beyerWorkExternalQuantum2020,deffnerQuantumWorkThermodynamic2016}. 
 {A similar construction was studied in Ref.~\cite{allahverdyanFluctuationsWorkQuantum2005} without providing the inequality~\eqref{eq:theorem}. The latter}
was first conjectured in Ref.~\cite{beyerMeasurementQuantumWork2023}. We prove it in App.~\ref{app:proof}  {based on a concavity result of Lieb~\cite{liebConvexTraceFunctions1973,ruskaiInequalitiesQuantumEntropy2002}}.

\paragraph*{Experimental relevance---}
Let us define the following quantity
\begin{align}
    \dFtilde = -\beta^{-1} \ln\left( \sum_i p_i e^{-\beta   {\wi}   }\right).
\end{align}
According to inequality \eqref{eq:theorem}, this is an upper bound to the true free energy difference and, by virtue of Jensen's inequality, it is a lower bound to the average work performed on the initial Gibbs state   {\(W_\text{avg} = \sum_i p_i \wi\)}  , i.e.,
\begin{align}
    W_\text{avg} \geq \dFtilde \geq \dF.
\end{align}
Both \(W_\text{avg} \) and \(\dFtilde\) are quantities that can be operationally determined in our scenario without knowledge of the system Hamiltonian, whereas the true free energy difference \(\dF\) remains inaccessible.
The average work \(W_\text{avg}\) has long been known to be an upper bound for \(\Delta F\). Our result shows that whenever the experimenter can gain some classical knowledge about the decomposition of the thermal state by measuring the environment, the quantity \(\dFtilde\) provides a tighter estimate of \(\dF\).

The bound \(\dFtilde\) depends on the unitary evolution \(U\) and the decomposition \(\mathcal{D}\) realized in the system. \(U\) and \(\mathcal{D}\) are hidden for the experimenter. However, by changing the protocol \(\mu_t\) (keeping \(\mu_0\) and \(\mu_T\) fixed) and by implementing different measurements \(\A\) on the environment, the experimenter can indirectly vary \(U\) and \(\mathcal{D}\), and try to minimize \(\dFtilde\).
In fact,  {\(\min_{U,\mathcal{D}} \dFtilde = \dF\)}.
Thus, in principle the correct \(\dF\) could be determined operationally. However, an experimenter will never know whether they have actually reached the true minimum. 

From Eq.~\eqref{eq:theorem}, it directly follows that the inequality is saturated if \(\mathcal{D}\) is the eigendecomposition of \(\rho_0\) and \([U^\dag H_T U, H_0] = 0\). 
This can be seen as a quasiclassical special case, since it does not involve quantum coherences between the energy eigenstates at the beginning and at the end of the protocol.
Thus, there is a clear quantum disadvantage for the operational determination of \(\dF\) from nonequilibrium work measurements. 

 {Importantly, the discrepancy between the true \(\dF\) and the measurable \(\dFtilde\) is a fundamental quantum effect. It should not be confused with the bias and uncertainty arising from limited measurement data in classical Jarzysnki experiments~\cite{talknerColloquiumStatisticalMechanics2020,shirtsEquilibriumFreeEnergies2003,jarzynskiRareEventsConvergence2006,dengDeformedJarzynskiEquality2017}, but persists even under the idealization of no statistical uncertainty for the measured work as assumed here.}

\paragraph*{Open systems---}
Analogous to the classical case, and in contrast to the TPM JE, our fluctuation theorem   {\eqref{eq:theorem} holds in the very same way for open systems, i.e.,} systems that partially thermalize during the driving.
We will describe the open system scenario in a stepwise way where the driving is interrupted by thermal damping steps. This is inspired by the original Ref.~\cite{jarzynskiEquilibriumFreeenergyDifferences1997}~Sec.~III.~D for the classical case. 

We again consider a time-continuous protocol with initial Hamiltonian \(H_0\) at time \(t_0\) and final Hamiltonian \(H_T\) at time \(T\). At times \(t_n\in \{t_1, \ldots, t_{N}=T\}\), the system undergoes a thermal damping step which is assumed to take no time. 
The instantaneous system Hamiltonians at these times \(t_n\) are denoted by \(H_n\). The damping is modelled by completely positive and trace-preserving maps \(\mathcal{K}_n\) that are Gibbs-preserving with respect to the  \(H_n\), i.e., \(\mathcal{K}_n[\rho_n] = \rho_n\), where \(\rho_n = \exp[-\beta H_n]/\tr \exp[-\beta H_n]\) (the system will, however, generally not be in that state at time \(t_n\)). As in the classical case, a detailed balance condition is not required~\cite{jarzynskiEquilibriumFreeenergyDifferences1997}.
Between the damping steps, the system evolution is governed by the unitary evolution due to the time-dependent system Hamiltonian \(H_{\mu_t}\).
The work measured in such a protocol is then
\begin{align}
\label{eq:thermalized-work}
     {\wi} = \sum_{n=1}^{N}\tr[H_n\rho_i^{(n)_-}] - \tr[H_{n-1}\rho_i^{(n-1)_+}],
\end{align}
where \(i\) denotes the label of the initial state in the decomposition \(\mathcal{D}\) as before, and the superscripts \((n)_-\) and \((n)_+\) denote the state of the system right before and right after the thermal damping at time \(t_n\), respectively.
The work values  {\(\wi\)} in Eq.~\eqref{eq:thermalized-work} also satisfy the inequality~\eqref{eq:theorem}, making the operational fluctuation theorem applicable to open quantum systems.
The proof is given in the Supplemental Material~\footnote{Supplemental material}. An extension to the time-continuous case is straightforward~\cite{beyerMeasurementQuantumWork2023}. 

\paragraph*{Conclusion---}
In this letter, we construct {{an operational quantum work measurement scenario that satisfies a quantum fluctuation relation for the free energy difference \(\dF\) with two crucial features: (i) the system Hamiltonian can remain unknown to the experimenter and (ii) the theorem applies to open systems as well.
While in the original classical setting both these properties hold, they cease to be fulfilled for the commonly adopted quantum TPM scheme.
We argue that in the TPM scenario, performing a Jarzynski experiment lacks predictive power regarding the determination of \(\dF\), as this quantity can be directly obtained from the known Hamiltonians.

We {{clarify}} that the operationality of the classical Jarzynski equality is based on the fact that \ref{it:1} a classical system is always in an objective microstate and that \ref{it:2} classical quantities can be measured in a single experimental run.
Both conditions do not hold for quantum systems, where states have to be objectified by measurements and physical quantities have to be measured as expectation values over many runs of the same experiment. 
Bearing this in mind, we construct a scenario that obtains objective data about the decomposition of the thermal state through measurements on its environment.
Conditioning the work measurement on this information leads to a {{useful quantum fluctuation theorem}} for the free energy difference \(\dF\).
Interestingly, our theorem comes in the form of an inequality. Equality is reached for a quasiclassical case with no energy coherences in the decomposition, which shows a general quantum disadvantage.

 {Our framework paves the way for further theoretical investigations. In particular, it is an open question whether a Crooks-type fluctuation theorem can be formulated in the same operational spirit.}
Experimental implementations of our scenario are readily feasible with current quantum technologies, and might even be simpler than the realization of the sequential projective measurements in the  {standard} TPM scheme.}

\begin{acknowledgments}
The authors would like to thank Kimmo Luoma for many fruitful discussions on this topic.
\end{acknowledgments}





\appendix
\section{External quantum work measurement}
\label{app:work-measurement}
 {Here, we describe a possible construction of the apparatus used in the main text. Details and variations can be found in Ref.~\cite{beyerMeasurementQuantumWork2023}.}
In the classical case, a crucial requirement for a successful external work measurement is the fact that the driving apparatus is at the same time also the device that measures the work (see Fig.~\ref{fig:molecule}).
Analogously, we define a quantum \emph{meter} \(M\) that interacts with the system of interest \(S\)~\cite{beyerWorkExternalQuantum2020,beyerMeasurementQuantumWork2023}.  {(See also Ref.~\cite{hanQuantumLimitationExperimental2024} for a different approach.)}

Let us consider a time-continuous family of normalized pure quantum states \(\ket{\mu_t}\), \(t \in [0,T]\), on \(M\).
We call \(\ket{\mu_t}\) the \emph{protocol} and define the projector and its time-derivative
\begin{align}
    \M_t = \kb{\mu_t}{\mu_t},\quad\Mdot_t = \frac{\partial }{\partial t } \M_t.  
\end{align}
The experimenter has all information about the meter \(M\) and predefines the protocol \(\ket{\mu_t}\).
Furthermore, we define a stroboscopic protocol by chopping the total time \(T\) into \(N\) pieces of length \(\Delta T = T/{{N}}\), and define
\begin{align}
\label{eq:Ms}
    \M_n = \M_{t = n \Delta t}, && \Mdot_n = \Mdot_{t=n \Delta t}.
\end{align}

The meter \(M\) interacts with the system \(S\) via a joint Hamiltonian \(\HSM\). The experimenter is ignorant about the details of \(S\) and about the Hamiltonian \(\HSM\). Nevertheless, they are in principle able to implement the experiment depicted in Fig.~\ref{fig:measured}.
At time \(t_n = n \Delta t\), the experimenter prepares their system in the pure state \(\M_n\). During a time step \(\Delta t\), the system interacts unitarily according to the joint Hamiltonian \(\HSM\) before it is measured by an observable \(\Omega_n\) and reprepared in state \(\M_{n+1}\) at time \(t_{n+1}\).
\begin{figure}[ht]
    \centering
    \includegraphics[width=\linewidth]{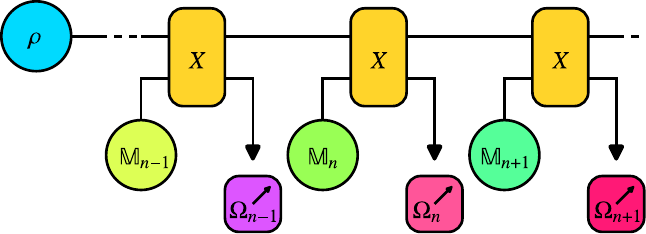}
    \caption{Measurement scheme that implements an effective unitary evolution on \(S\) and measures the performed work. In each step, the meter is prepared in a pure state \(\M_n\) according to a stroboscopic version of a time-continuous protocol \(\ket{\mu_t}\). After a short interaction \(X = \exp[-i \Delta t \HSM]\), the work performed on \(S\) in this step can be measured as an observable \(\Omega_n\) on the meter \(M\).}
    \label{fig:measured}
\end{figure}

\subsection{Hamiltonian evolution in \(S\)}

For short time steps \(\Delta t\), this scheme effectively implements a unitary evolution in the system \(S\).
At time \(t_n\), the system is in the (unknown) state \(\rho_{n}\). Neglecting terms of order \(O(\Delta t^2)\), the evolution in the next time step is described by
\begin{align}
\label{eq:strob-approx}
    \rho_{n+1} =& {{\tr_M}}\{e^{-i \HSM \Delta t} (\rho_{n} \otimes \M_n) e^{i \HSM \Delta t}\}  \notag \\    
    =&\rho_{n} -i \Delta t \tr_M\{[\HSM,\rho_{n} \otimes \M_n] \} \notag \\
     =&\rho_{n} -i \Delta t [\bra{\mu_n}\HSM\ket{\mu_n},\rho_{n}].
\end{align}
In the limit \(N \to \infty\), the stroboscopic evolution becomes continuous and we get
\begin{align}
    \dot \rho(t) &= \lim_{\Delta t \to 0} \frac{\rho(t + \Delta t) - \rho(t)}{\Delta t} \notag \\
    &= -i [\bra{\mu_t}\HSM\ket{\mu_t},\rho(t)].
\end{align}
Thus, for sufficiently short \(\Delta t\), the stroboscopic experiment implements a unitary evolution on system \(S\) which is governed by the time-dependent Hamiltonian 
\begin{align}
    H_{\mu_t} = \bra{\mu_t} \HSM \ket{\mu_t}.
\end{align}
The subscript indicates that the effective Hamiltonian in \(S\) depends on the protocol of quantum states \(\ket{\mu_t}\) the experimenter chooses to implement. 

Some remarks are in order here. 
Since we have discarded all terms of order \(O(\Delta t^2)\) in Eq.~\eqref{eq:strob-approx}, the system does not entangle with the meter (otherwise the effective dynamics in \(S\) would not be unitary). 
This is in contrast to collision models for open system dynamics where a second-order expansion is used together with further arguments of coarse graining in order to obtain a non-unitary dynamics in the system~\cite{ciccarelloQuantumCollisionModels2022}. 
The regime described here is sometimes called the dynamical quantum Zeno effect~\cite{facchiQuantumZenoDynamics2008,pascazioDynamicalQuantumZeno1994}. In fact, choosing a trivial protocol \(\ket{\mu_t} = \ket{\mu_0}\), the experimenter could simply measure projectively with high frequency in order to freeze the state in \(\ket{\mu_0}\) as in the common quantum Zeno effect. 
Crucially, a Zeno measurement on subsystem \(M\) does not freeze the dynamics in subsystem \(S\)~\cite{facchiQuantumZenoEffect2003}.

\subsection{Work measurement}
So far, we have shown that the experiment in Fig.~\ref{fig:measured} effectively implements a unitary evolution on \(S\) that is governed by a Hamiltonian that depends on the time-dependent protocol \(\ket{\mu_t}\) the experimenter implements, but is otherwise unknown. 

Since the system \(S\) is effectively closed, the work performed on it is equal to the energy difference with respect to the effective Hamiltonian. For an initial state \(\rho\), we expect
\begin{align}
\label{eq:average-work-app}
      {w_\qu[\rho]} = \tr[\rho (U^\dag H_T U-H_0)],
\end{align}
where
\begin{align}
    H_0   = \bra{\mu_0} \HSM \ket{\mu_0}, &&
    H_T   = \bra{\mu_T} \HSM \ket{\mu_T}.
\end{align}
We now show that the quantity \(w_\qu[\rho]\) can be measured on the meter \(M\) with arbitrary precision if suitable observables \(\Omega_n\) are implemented (see Fig.~\ref{fig:measured}).
  
In the \(n\)th step, the meter is prepared in state \(\M_n\). We denote the state of the meter after the interaction with the system \(S\) by
\begin{align}
    {\chi_{n}} = \M_n - i \Delta t \tr_S \{ [\HSM,\rho_{n} \otimes \M_n] \},
\end{align}
where {{we}} have again neglected terms of order \(O(\Delta t^2)\).
A suitable work observable on the meter \(M\) after its interaction is given by~\cite{beyerMeasurementQuantumWork2023,beyerWorkExternalQuantum2020}
\begin{align}
    \Omega_n = -i[\Mdot_n, \M_n],
\end{align}
with \(\Mdot\) from Eq.~\eqref{eq:Ms}.
Importantly, since the experimenter knows the implemented protocol \(\ket{\mu_t}\), they can also construct the observable \(\Omega\).
Its expectation value reads
\begin{align}
    \langle \Omega_n \rangle =& \tr\left\{ \Omega_n \chi_{n} \right\}  \\ =& - i \tr\left\{[\Mdot_n, \M_n]\M_n\right\} \notag\\
    &- \Delta t \tr\left\{[\Mdot_n, \M_n] [\HSM,\rho_{n} \otimes \M_n]  \right\}. \notag
\end{align}
The first term vanishes. Using the identities
\begin{align}
    \M\Mdot + \Mdot\M = \partial_t\M^2 = \Mdot,
\end{align}
and
\begin{align}
    \M\Mdot\M = \M(\M\Mdot + \Mdot\M)\M = 2 \M\Mdot\M = 0,
\end{align}
we find for the second term
\begin{align}
    \langle \Omega_n \rangle = \Delta t \tr\left\{  \rho_{n} \Mdot_n \HSM \right\}.
\end{align}
As required, this is exactly the energy change of the system \(S\) in step \(n\) since
\begin{align}
\label{eq:energy-difference}
    E_{n+1} &= \tr\left\{\rho_{n+1} H_{{\mu_{n+1}}} \right\} = \tr\left\{\rho_{n+1} \M_{n+1} \HSM  \right\}\notag\\
    &=\tr\left\{\left(\rho_{n} - i \Delta t[\M_n \HSM, \rho_{n}] \right) \ldots \right. \notag\\
    &\quad\quad\quad\quad\quad\quad\quad \left. \ldots\left(\M_n+ \Delta t \Mdot_n \right) \HSM \right\} \notag \\
    &= E_{n} + \Delta t \tr\left\{ \rho_{n} \Mdot_n \HSM \right\},
\end{align}
which is the discretized version of the fact that 
\begin{align}
 \dot E \, \diff t = \tr[\rho \dot H] \diff t
\end{align}
in closed systems.
For a sufficiently small time step \(\Delta t\) the sum over the work measured in the single steps approximates the expected total average work in Eq.~\eqref{eq:average-work-app}, i.e.,
\begin{align}
     \sum_n \langle \Omega_n \rangle \approx   {w_\qu[\rho]}.
\end{align}
Mathematically, the approximation becomes exact by replacing the sum by an integral in the limit of \mbox{\(\Delta t \to 0\)}. 
However, this limit is practically infeasible as the variance of the \(\Omega_n\) diverges with \(\Delta t \to 0\) (see Ref.~\cite{beyerMeasurementQuantumWork2023}~Ch.~5.2). 
This is consistent with the fact that the time-continuous limit leads to a perfect unitary evolution in \(S\) and, according to the principle of no information without disturbance, the measurement must not yield any information about the system in this case. 
This resembles the fact that also in the classical experiment the spring of the meter in Fig.~\ref{fig:molecule} has to move in order to measure the force \(F\), thereby consuming some of the work. However, in a classical case the effect can in principle be made arbitrarily small without increasing the variance of the measurement, whereas in the quantum case a better and better approximation of the perfect unitary dynamics comes at the expense of more and more experimental runs needed to estimate the expectation values of the observables \(\Omega_n\).
A scheme to partially overcome this problem by implementing the continuous scheme but coarse-graining the work observable has been given in Ref.~\cite{beyerMeasurementQuantumWork2023}~Ch.~5.6.
Conceptually important for this letter is only the fact that the work in Eq.~\eqref{eq:expected-work} can in principle be measured to arbitrary finite precision.

\section{Proof of the fluctuation theorem (closed system)}
\label{app:proof}

We define \(\rhoA = \rho_0\) and \(\rhoB = Z_T^{-1} \exp[{-\beta U^\dag H_T U}] \).  {Using this and Eq.~\eqref{work_i} we}  rewrite Eq.~\eqref{eq:theorem} as 
\begin{align}
\label{eq:JIE-ln}
\sum_i p_i e^{\tr[\rho_i(\ln\rhoB-\ln\rhoA)]} \leq 1 ,\; \text{with}\; \sum_i p_i \rho_i = \rhoA,
\end{align}
 {with \(\mathcal{D} = \{(p_i,\rho_i)\}\) being an arbitrary decomposition of the initial Gibbs state.} 

The proof of (\ref{eq:JIE-ln}) starts with what is known as the Peierls-Bogoliubov inequality~\cite{wehrlGeneralPropertiesEntropy1978}
\begin{align}\label{PBIE}
    \tr[e^{A}] e^{\langle B \rangle} = \tr[e^{A + \langle B \rangle}] \le \tr\left[e^{A + B}\right],
\end{align}
 where $\langle B\rangle = \tr[e^A B]/\tr[e^A]$.
Here, we choose $B=\ln \rho_B - \ln \rho_A$ and $e^A = \rho_i$ in (\ref{PBIE}) so that
\begin{align}\label{1ststep}
e^{\tr[\rho_i(\ln \rho_B - \ln \rho_A)]}\le \tr\left[e^{\ln \rho_i + \ln \rho_B - \ln \rho_A}\right].
\end{align}
Furthermore, we employ a concavity result for operator functions due to Lieb \cite{liebConvexTraceFunctions1973,ruskaiInequalitiesQuantumEntropy2002}:
\begin{equation}\label{Lieboperatorfunction0}
    A\rightarrow F_L[A]\equiv \tr[e^{\ln A + L}]
\end{equation}
is concave in positive $A$ for fixed Hermitian $L$.
Thus,
\begin{align}
     &\sum_i p_i e^{\tr[\rho_i(\ln \rho_B - \ln \rho_A)]} \le \sum_i p_i \tr\left[e^{\ln \rho_i + \ln \rho_B - \ln \rho_A}\right]  \notag \\
 & \le \tr\left[e^{\ln(\sum_i p_i \rho_i) + \ln \rho_B - \ln \rho_A}\right] = \tr[\rho_B] = 1. \quad \blacksquare
\end{align}

\bibliography{bib}

\clearpage

\title{Supplemental Material: Operational Work Fluctuation Theorem for Open Quantum Systems}

\maketitle

\setcounter{page}{1}
\renewcommand{\theequation}{S\arabic{equation}}
\setcounter{equation}{0}

\section*{Proof for open quantum systems}
We start by proving the open system case of Eq.~\eqref{eq:theorem} with the work  {\(\wi\)} given in Eq.~\eqref{eq:thermalized-work} for a single intermediate thermal damping step, before we generalize it to multiple steps.

The scenario is described by an initial Hamiltonian \(H_A = H_0\), an intermediate one \(H_B\) at time \(t=t_B\), and a final one \(H_C = H_T\), with the corresponding Gibbs states 
\begin{align}
    \rho_X = \frac{\exp[-\beta H_X]}{\tr\exp[-\beta H_X] }, \quad X\in\{A,B,C\}.
\end{align}
Without loss of generality, we assume that the unitary evolution before and after the intermediate damping is given by the identity \(U = \id\), since we can always absorb any nontrivial unitary dynamics into the Hamiltonians. 

The system undergoes a thermal damping step at time \(t_B\) described by a quantum channel \(\mathcal{K}\) that is Gibbs-preserving with respect to the instantaneous Hamiltonian, i.e., 
\begin{align}
\label{eq:fixpoint}
    \K[\rho_B] = \rho_B.
\end{align}
The damping is assumed to take no time. Before and after the intermediate damping, the work can be measured as in the closed system case, i.e.,
\begin{align}
      {\wi} = \tr\left[\rho_i(H_B - H_A)\right] + \tr\left[\K[\rho_i](H_C - H_B)\right].
\end{align}
Defining
\begin{align}
\label{eq:w-tilde}
    \tilde{w}_\qu[\rho_i] = \tr[\rho_i(\ln \rho_B - \ln \rho_A)]+ \tr[\K[\rho_i](\ln \rho_C - \ln \rho_B)],
\end{align}
the inequality we have to prove can now be written as
\begin{align}
\label{eq:JIE_thermal}
    \sum_i p_i e^{ {\tilde{w}_\qu[\rho_i]}} \leq 1,
\end{align}
with the ensemble condition for the initial Gibbs state
\begin{align}
\label{eq:ensemble_A}
    \rho_0 = \rho_A = \sum_i p_i \rho_i.
\end{align}

The thermalizing map $\K$ has a Stinespring dilation (including an ancillary environment $E$) of the form
\begin{align}\label{kraus}
      \K[\rho] = \tr_E[U_{SE}\,(\rho\otimes \varepsilon) \,U^\dagger_{SE}].
\end{align}
with some fictitious environmental initial state $\varepsilon$. (This state must not be confused with the actual state of a physical environment. In particular, it can always be chosen to be pure.)
We aim to rewrite the two traces in Eq. (\ref{eq:w-tilde}), employing these environmental degrees of freedom. As an abbreviation, we will use the notation
\begin{align}\label{propagated}
    \rho_{xE} = U_{SE}\,(\rho_x\otimes \varepsilon) \,U^\dagger_{SE}, && x=i,A,B,
\end{align}
for the full, likely correlated, unitarily evolved system-environment state such that, e.g.
\begin{align}\label{shorthand}
    \K[\rho_i]=\tr_E \rho_{iE}.
\end{align}

For the first trace in Eq.~(\ref{eq:w-tilde}), we start by simply extending the system states by the initial environment state \(\varepsilon\) of the Stinespring dilation (\ref{propagated}) to a sytem-environment state: $\rho_x \rightarrow \rho_x\otimes\varepsilon$ with $x=i,A,B$ and find
\begin{align}\label{1sttr}\nonumber
    & \tr_S[\rho_i(\ln \rho_B - \ln \rho_A)]  \\ \nonumber
    & = \tr_{SE}\left[(\rho_i\otimes\varepsilon)\left(\ln (\rho_B \otimes\varepsilon)- \ln (\rho_A \otimes\varepsilon)\right)\right] \\
    &= \tr_{SE}[\rho_{iE}(\ln\rho_{BE}-\ln\rho_{AE})].
\end{align}
For the last line we used the unitary invariance of the $SE$-trace under $U_{SE}$ to replace the product initial states by their time-evolved (correlated) versions from Eq.~(\ref{propagated}).

The second trace in Eq.~(\ref{eq:w-tilde}) is rewritten somewhat similarly, using Eq.~(\ref{shorthand}) to replace the thermalizing map by an environmental trace:
\begin{align}\label{2ndtr}\nonumber
      &\tr_S[\K[\rho_i](\ln \rho_C - \ln \rho_B)] \\ \nonumber
            &=\tr_S[(\tr_E\rho_{iE})(\ln \rho_C - \ln \rho_B)]  \\
              &= \tr_{SE}[\rho_{iE}(\ln \rho_{C1} - \ln \rho_{B1})].
\end{align}
Here we have extended the arguments of the logarithms trivially by a fully mixed environment state to give an $SE$-operator,
\begin{align}\label{triviallift}
    \rho_x \rightarrow \rho_x\otimes \frac{\id_E}{d_E} \eqcolon \rho_{x1},
\end{align}
where $d_E$ is the dimension of the environmental Hilbert space.
Note that (despite the normalization factor)
$(\ln\rho_C - \ln\rho_B)\otimes \id_E = \ln\rho_{C1}-\ln\rho_{B1}$, as used to obtain the last line in Eq.~(\ref{2ndtr}).

We now replace and combine the two traces in the open quantum work inequality Eq.~(\ref{eq:JIE_thermal}) in terms of the Stinespring lifts in Eqns.~(\ref{1sttr},\ref{2ndtr},\ref{triviallift}),
involving environmental degrees of freedom.
Thus, our open quantum work inequality (\ref{eq:JIE_thermal}) takes the equivalent form
\begin{align}
\label{eq:JIE_thermal_lifted} \nonumber
& \sum_i p_i e^{\tr_S[\rho_i(\ln \rho_B - \ln \rho_A)]+ \tr_S[\K[\rho_i](\ln \rho_C - \ln \rho_B)]} = \\
 &    \sum_i p_i e^{\tr_{SE}[\rho_{iE}(\ln \rho_{BE} - \ln \rho_{AE} + \ln \rho_{C1} - \ln \rho_{B1})]} \leq 1,
\end{align}
which we are going to prove in what follows.

First, we take the two very same steps as in the proof of the closed case in App.~\ref{app:proof}: The Peierls-Bogoliubov inequality similar to Eq.~(\ref{1ststep}), and Lieb's concavity theorem as in Eq.~(\ref{Lieboperatorfunction0}), and,  moreover, employ the fact that $\sum_i p_i \rho_{iE} = \rho_{AE}$ from Eq.~(\ref{eq:ensemble_A}) and Eq.~(\ref{propagated}). From these steps we find for the left-hand side of Eq.~(\ref{eq:JIE_thermal_lifted}) the upper bound
\begin{align}
\label{eq:JIE_thermal_lifted_2} \nonumber
   &\sum_i p_i e^{\tr_{SE}[\rho_{iE}(\ln \rho_{BE} - \ln \rho_{AE} + \ln \rho_{C1} - \ln \rho_{B1})]}  \\
   & \leq \tr_{SE}[e^{\ln \rho_{BE} + \ln \rho_{C1} - \ln \rho_{B1}}].
\end{align}
To proceed, we now apply the 
Lieb-Golden-Thompson inequality \cite{liebConvexTraceFunctions1973,ruskaiInequalitiesQuantumEntropy2002}
\begin{align}\label{LGTIE}
   &\tr\left[e^{\ln T + \ln R - \ln S}\right] \notag \\
   &\le \tr\left[\int_0^\infty \diff u \, R (S+u\id)^{-1} T (S+u\id)^{-1}\right].
\end{align}
We choose $T=\rho_{BE}, R=\rho_{C1}$, and $S=\rho_{B1}$ and find the desired upper bound for the right-hand side of Eq.~(\ref{eq:JIE_thermal_lifted_2}),
\begin{align}
\label{eq:JIE_thermal_lifted_3} \nonumber
   & \tr_{SE}[e^{\ln \rho_{BE} + \ln \rho_{C1} - \ln \rho_{B1}}]  \\ \nonumber
   & \leq \tr_{SE}\!\left[\int_0^\infty \!\!\diff u \,  \rho_{C1} (\rho_{B1}\!+\!u\id_{SE})^{-1} \!\rho_{BE} (\rho_{B1}\!+\!u\id_{SE})^{-1}\right]  \\ \nonumber
   & = \tr_S\!\left[\int_0^\infty \!\!\diff u' \, \rho_{C} (\rho_{B}\!+\!u'\id_{S})^{-1} \!\tr_E[\rho_{BE}] (\rho_{B}\!+\!u'\id_{S})^{-1}\right]  \\ \nonumber
 & = \tr_S\!\left[\int_0^\infty \diff u' \, \rho_{C} (\rho_{B}+u'\id_{S})^{-1} \rho_{B} (\rho_{B}+u'\id_{S})^{-1}\right]  \\ 
 & = \tr_S\rho_C = 1.
\end{align}
Here, the bound from the first to the second line is simply the Lieb-Golden-Thompson inequality (\ref{LGTIE}). The third line follows from the second by noting that all three operators $\rho_{C1}, \rho_{B1}$, and $\id_{SE}$ have a trivial environmental part [see Eq.~(\ref{triviallift})], such that the environmental trace needs to be taken around the state $\rho_{BE}$, only. In order to get the normalizations right [see again Eq.~(\ref{triviallift})], it is necessary to scale the integration variable by the dimension of the environmental Hilbert space $u\rightarrow u'=d_E u$. 
The fourth line follows from the third due to the properties of the thermalizing map and its fixed point: from Eqns.~(\ref{eq:fixpoint}, \ref{shorthand}) we see $\tr_E[\rho_{BE}]=\K[\rho_B]=\rho_B$. Finally, using a spectral decomposition of $\rho_B$, it is clear that $\int_0^\infty \diff u' \, (\rho_{B}+u'\id_{S})^{-1} \rho_{B} (\rho_{B}+u'\id_{S})^{-1} = \id_S $ which completes the proof for a single damping step. \hfill \(\blacksquare\)

\subsection*{Multiple steps}
It only remains to show that the inequality also holds for any subsequent step. 
Adding a damping step with respect to Hamiltonian \(H_C\) and a final Hamiltonian \(H_D\), the exponent in Eq.~\eqref{eq:JIE_thermal} reads
\begin{align}
\label{eq:work-multiple-steps}
      {\tilde{w}_\qu[\rho_i]} =& \tr[\rho_i(\ln \rho_B-\ln\rho_A)]\notag\\ &+ \tr[K_B[\rho_i](\ln \rho_C - \ln \rho_B)]\notag\\ &+ \tr[K_C\circ K_B[\rho_i](\ln\rho_D -\ln\rho_C)].
\end{align}
Along the same lines as before in Eq.~\eqref{propagated}, we can extend each state to a now tripartite state on the system \(S\) and two ancillary environments \(E\) and \(F\) that implement a Stinespring dilation of the first and the second damping step, respectively.
We redefine
\begin{align}
    \rho_{xEF} &= U_{SF}U_{SE}(\rho_x\otimes \varepsilon \otimes \varphi) U^\dagger_{SE}U_{SF}^\dag, \notag\\
    \rho_{x1F} &= U_{SF}\left(\rho_x\otimes \frac{\id_E}{d_E} \otimes \varphi\right) U_{SF}^\dag, \notag\\
    \rho_{x11} &=\rho_x\otimes \frac{\id_E}{d_E} \otimes \frac{\id_F}{d_F}, \notag \\
    \rho_{xF} &= U_{SF}\left(\rho_x\otimes \varphi\right) U_{SF}^\dag, \notag \\
    \rho_{x1} &= U_{SF}\left(\rho_x\otimes \frac{\id_F}{d_F}\right) U_{SF}^\dag,
\end{align}
with \(x = i,A,B,C,D\) and rewrite Eq.~\eqref{eq:work-multiple-steps} as
\begin{align}
     {\tilde{w}_\qu[\rho_i]} =& \tr[\rho_{iEF}(\ln \rho_{BEF} - \ln\rho_{AEF})] \notag \\
    &+ \tr[\rho_{iEF}(\ln \rho_{C1F} - \ln\rho_{B1F})] \notag\\
    &+\tr[\rho_{iEF}(\ln \rho_{D11} -\ln\rho_{C11}].
\end{align}
Using the same reasoning as for Eq.~\eqref{eq:JIE_thermal_lifted_2} we arrive at
\begin{align}
    \sum_i &p_i e^{  {\tilde{w}_\qu[\rho_i]}} \!\leq\! \tr[e^{\ln \rho_{BEF} + \ln \rho_{C1F} - \ln\rho_{B1F} + \ln \rho_{D11} -\ln\rho_{C11}  }].
\end{align}
For the Lieb-Golden-Thomson inequality in Eq.~\eqref{LGTIE} we now choose \(T = \rho_{BEF}\), \(S=\rho_{B1F}\), and \(\ln R = (\ln \rho_{C1F} + \ln\rho_{D11} - \ln \rho_{C11})\).
We then get
\begin{align}
\label{eq:thermal-LGT}
    &\tr_{SEF}[e^{\ln \rho_{BEF} + \ln \rho_{C1F} - \ln\rho_{B1F} + \ln \rho_{D11} -\ln\rho_{C11} }] \notag \\
    &\leq \tr_{SF}[ R_{SF} \int \diff u' (\rho_{BF} + u'\id)^{-1} \tr_E[\rho_{BEF}] (\rho_{BF} + u'\id)^{-1} \notag \\
    &=\tr_{SF}[ R_{SF} \int \diff u' (\rho_{BF} + u'\id)^{-1} \rho_{BF} (\rho_{BF} + u'\id)^{-1} \notag \\
    &=\tr_{SF}[ R_{SF} \int \diff u' (\rho_{BF} + u'\id)^{-1} \rho_{BF} (\rho_{BF} + u'\id)^{-1} \notag \\
    &= \tr_{SF}[ R_{SF} \id] \notag \\
    &= \tr_{SF}[ e^{\ln \rho_{CF}+ \ln\rho_{D1} - \ln \rho_{C1} }],
\end{align}
where in the second line we have used that
\begin{align}
    \ln (\rho \otimes \frac{\id}{d}) = \ln(\rho) \otimes \id - \ln d, 
\end{align}
and therefore
\begin{align}
    R &= e^{\ln \rho_{C1F} + \ln\rho_{D11} - \ln \rho_{C11} } \notag\\
    &= e^{\ln \rho_{CF} + \ln\rho_{D1} - \ln \rho_{C1} } \otimes \frac{\id_E}{d_E} \notag \\
    &\eqcolon R_{SF} \otimes \frac{\id_E}{d_E}.
\end{align}
From Eq.~\eqref{eq:thermal-LGT} we can continue as in the case with only a single damping step in Eq.~\eqref{eq:JIE_thermal_lifted_3}. 
Any additional damping steps can be reduced in the same way. \hfill \(\blacksquare\)

\end{document}